\documentclass[aps,pre,twocolumn]{revtex4}
\usepackage{graphics}
\usepackage[pdftex]{graphicx}
\usepackage{times}
\usepackage{multirow}
\usepackage[pdftex]{hyperref}
\hypersetup{colorlinks=false,pdfborder=0}
\usepackage{color}

\begin{document}

\title{Self-organization of progress across the century of physics}

\author{Matja{\v z} Perc}
\thanks{Electronic address: \href{mailto:matjaz.perc@uni-mb.si}{\textcolor{blue}{matjaz.perc@uni-mb.si}}}
\affiliation{Faculty of Natural Sciences and Mathematics, University of Maribor, Koro{\v s}ka cesta 160, SI-2000 Maribor, Slovenia}

\begin{abstract}
\noindent We make use of information provided in the titles and abstracts of over half a million publications that were published by the American Physical Society during the past 119 years. By identifying all unique words and phrases and determining their monthly usage patterns, we obtain quantifiable insights into the trends of physics discovery from the end of the 19th century to today. We show that the magnitudes of upward and downward trends yield heavy-tailed distributions, and that their emergence is due to the Matthew effect. This indicates that both the rise and fall of scientific paradigms is driven by robust principles of self-organization. Data also confirm that periods of war decelerate scientific progress, and that the later is very much subject to globalization.
\end{abstract}

\maketitle

\noindent The 20th century is often referred to as the century of physics \cite{timeline}. From x-rays to the semiconductor industry, the human society today would be very different were it not for the progress made in physics laboratories around the World. And while amid the economic woes the budget for science is being cut down relentlessly \cite{clery_s10, eds_np12}, it seems now more than ever the need is there to remind the policy makers of this fact. Although to the layman the progress made on an individual level may appear to be puny and even needless, the history teaches us that collectively the physics definitively delivers. It is therefore of interest to understand how the progress made so far came to be, and how to best maintain it in the future. Should there be overarching authorities that dictate which scientific challenges to address and prioritize, or should we rely on the spontaneous emergence of progress?

We know, for example, that the acquisition of citations \cite{redner_pt05} as well as the acquisition of collaborators \cite{newman_pre01} are subject to preferential attachment. These two processes are neither regulated nor imposed. They are perpetuated by scientific excellence and individual choice. In fact, Barab{\'a}si and Albert \cite{barabasi_s99} have shown that preferential attachment and growth give rise to robust principles of self-organization that culminate in the emergence of scaling. Preferential attachment can be considered synonymous to the Matthew effect, which sociologist Robert K. Merton \cite{merton_sci68} coined based on the writings in the Gospel of St. Matthew for explaining discrepancies in recognition received by eminent scientists and unknown researchers for similar work. Derek J. de Solla Price \cite{price_sci65} observed the same phenomenon when studying the network of citations between scientific papers, and most recently also the longevity of careers has been found driven by the Matthew effect \cite{petersen_pnas11}.

\begin{figure}
\centering{\includegraphics[width = 8.5cm]{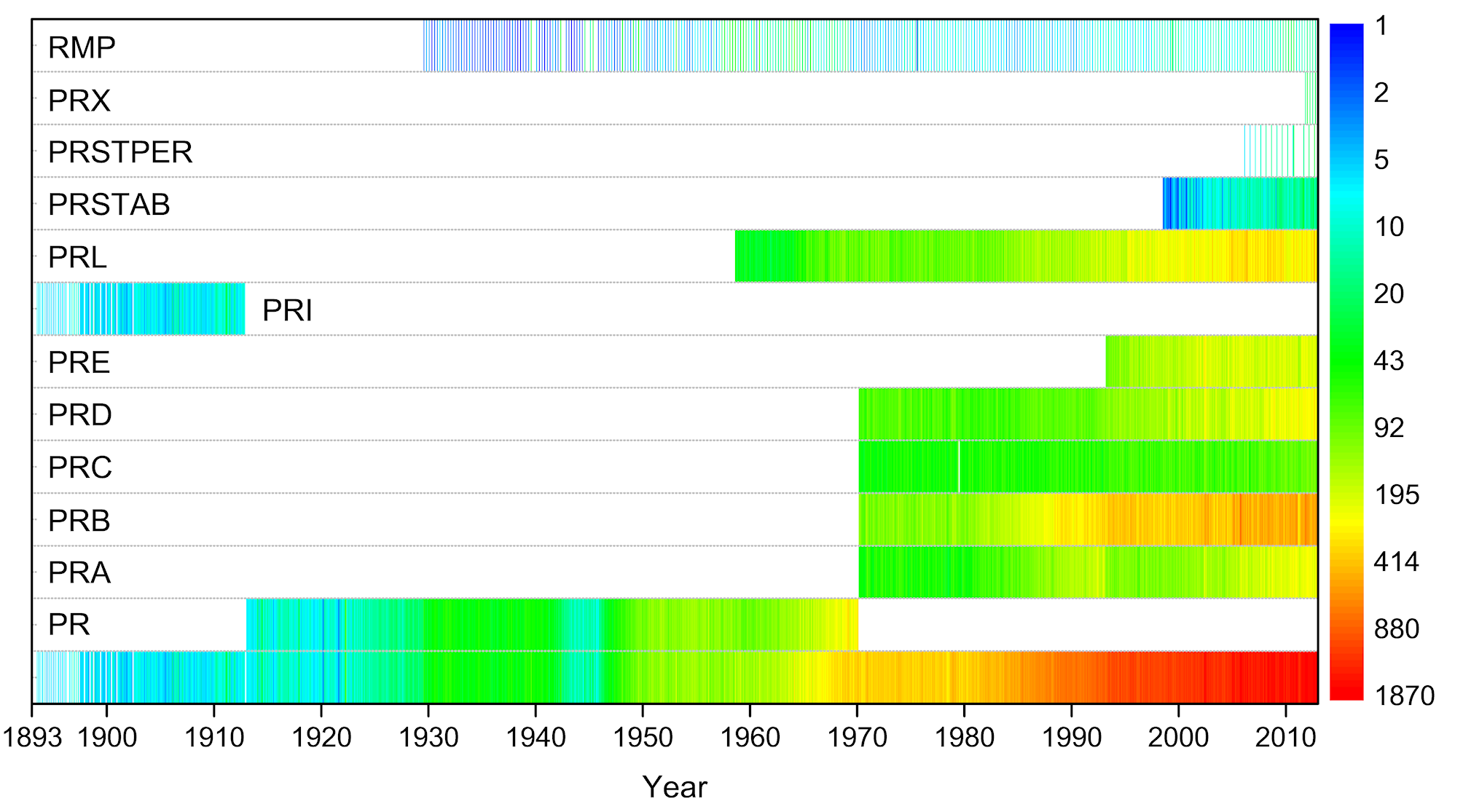}}
\caption{Publishing timeline of the Physical Review. Color encodes the number of publications per month for each particular journal (the color scale is logarithmic). Bottom most row depicts the overall output, corresponding to the sum of publications across all the journals. Volume 1, Issue 1 of Physical Review (Series I) (PRI) was published in July 1893. It consisted of five articles and two notes. Today the overall output hovers comfortably over 1600 publications per month, with the maximum being reach in June 2012, with 1870 publications. Physical Review B (PRB) has the largest number of publications per month, the record being 772 publications during July 2005. It can also be observed when certain journals where retired or introduced. Abbreviations of journal names are those commonly used by the American Physical Society.}
\label{data}
\end{figure}

Motivated by the existing reports of the Matthew effect in science, we explore whether the trends of scientific discovery are also subject to the same principles of self-organization. We make use of the titles and abstracts of over half a million publications of the Physical Review that were published between July 1893 and October 2012, and we infer the trends by adopting the methodology of culturomics \cite{michel_s11}. Our approach is thus purely data-driven \cite{evans_s11}, in line with substantial interdisciplinary research efforts that are currently aimed at obtaining quantitative insights into the social and natural sciences in general \cite{lazer_s09, barabasi_np12}, but also into sports \cite{radicchi_pone12}, drug discovery \cite{nussinov_tps11, csermely_pr13}, finance \cite{preis2012quantifying}, and scientific production \cite{radicchi_scirep12, pan_sr12, mazloumian_scirep13} in particular.

\section*{Results}

\begin{figure*}
\centering{\includegraphics[width = 11.6cm]{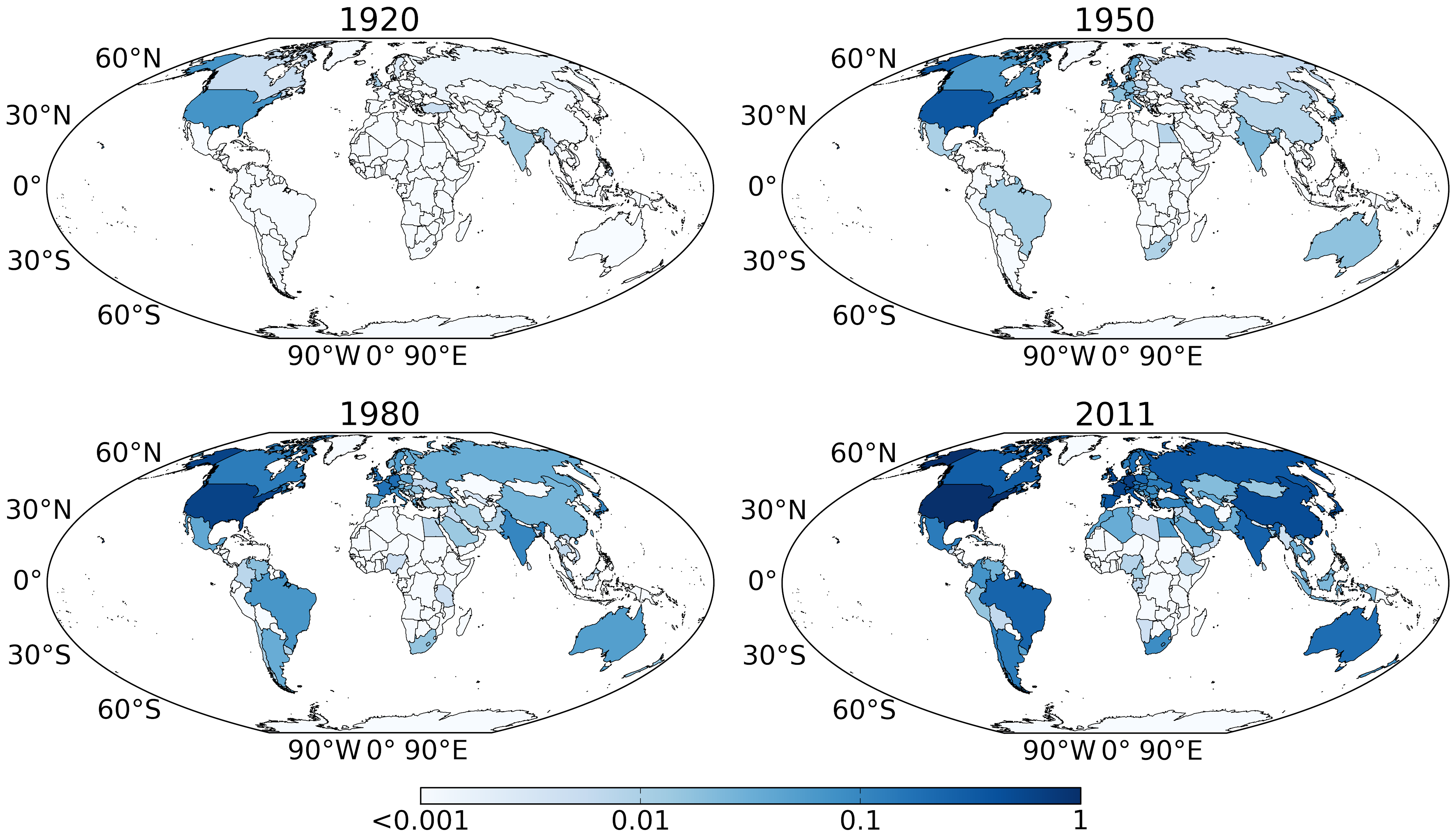}}
\caption{Countries that contribute to research that is published in the Physical Review. Color encodes the average monthly productivity of a country during each displayed year, as evidenced by the affiliations, normalized by the average monthly output of the US during 2011 (equalling $\approx 565$ publications per month -- a maximum). All affiliations were used, and in case more than one country was involved on a given publication, all received equal credit. A 12 month moving average was applied prior to calculating the average monthly production for each country. Note that the color scale is logarithmic. Displayed are World maps for four representative years, while the full geographical timeline can be viewed at \href{http://www.youtube.com/watch?v=0Xeysi-EfZs}{\textcolor{blue}{youtube.com/watch?v=0Xeysi-EfZs}}. We have used publicly available resources (\href{http://www.wikipedia.org/}{\textcolor{blue}{wikipedia.org}} and \href{http://maps.google.com/}{\textcolor{blue}{maps.google.com}}) to geocode the affiliations at the country level, as well as to disambiguate them in case of name variations, typos or name changes during the time period of study. Maps were produced with \href{http://matplotlib.org/basemap/}{\textcolor{blue}{matplotlib.org/basemap}} \cite{matplotlib}.}
\label{geo}
\end{figure*}

\noindent The timeline of publications for different journals and overall is presented in Fig.~\ref{data}. It can be observed that the overall output (bottom most color stripe) increases steadily over time. An obvious exception is the World War II period, during which the production dropped almost an order of magnitude, from nearly 100 publications per month before and after the war to below 10 during the war. This confirms, not surprisingly, that periods of war decelerate scientific progress or at least very much hinder the dissemination of new knowledge.

By geocoding the affiliations, it is also possible to infer where the physics published in the Physical Review has been coming from. As can be inferred from Fig.~\ref{geo} and the accompanying video, the formative years of Physical Review were dominated by the US, with relatively rare contributions coming from the UK, Germany, France and India. During the World Wars I and II large contingents of the World went silent (see the video referred to in the caption of Fig.~\ref{geo}), and it was only during the 1950s and 60s that the US centrism begun fading. The collapse of the Soviet Union, the fall of the Berlin Wall, and the related changes in World order during the 1980s and 90s contributed significantly to the globalization, so that today countries like China, Russia, Canada, Japan, Australia, as well as large regions of Europe and South America all contribute markedly to physics research that is published in the Physical Review. Countries that are still exempt are from Central Africa and Sahara. Ranking the countries according to their overall average monthly production during the last 20 years yields US, Germany, France, UK, Japan, Italy, China, Russia, Spain, Canada and Netherlands, while per capita yields Switzerland, Israel, Denmark, Sweden, Slovenia, Finland, Germany, Netherlands, France and Austria as the top 10, respectively. These results are in good agreement with more comprehensive rankings that were recently published based on World citation and collaboration networks over many different fields of research \cite{pan_sr12}. Our goal here is solely to provide a general overview of the geographical origin of the data, and so we proceed with the core analysis of trends of physics discovery.

\begin{figure}[b]
\centering{\includegraphics[width = 8.5cm]{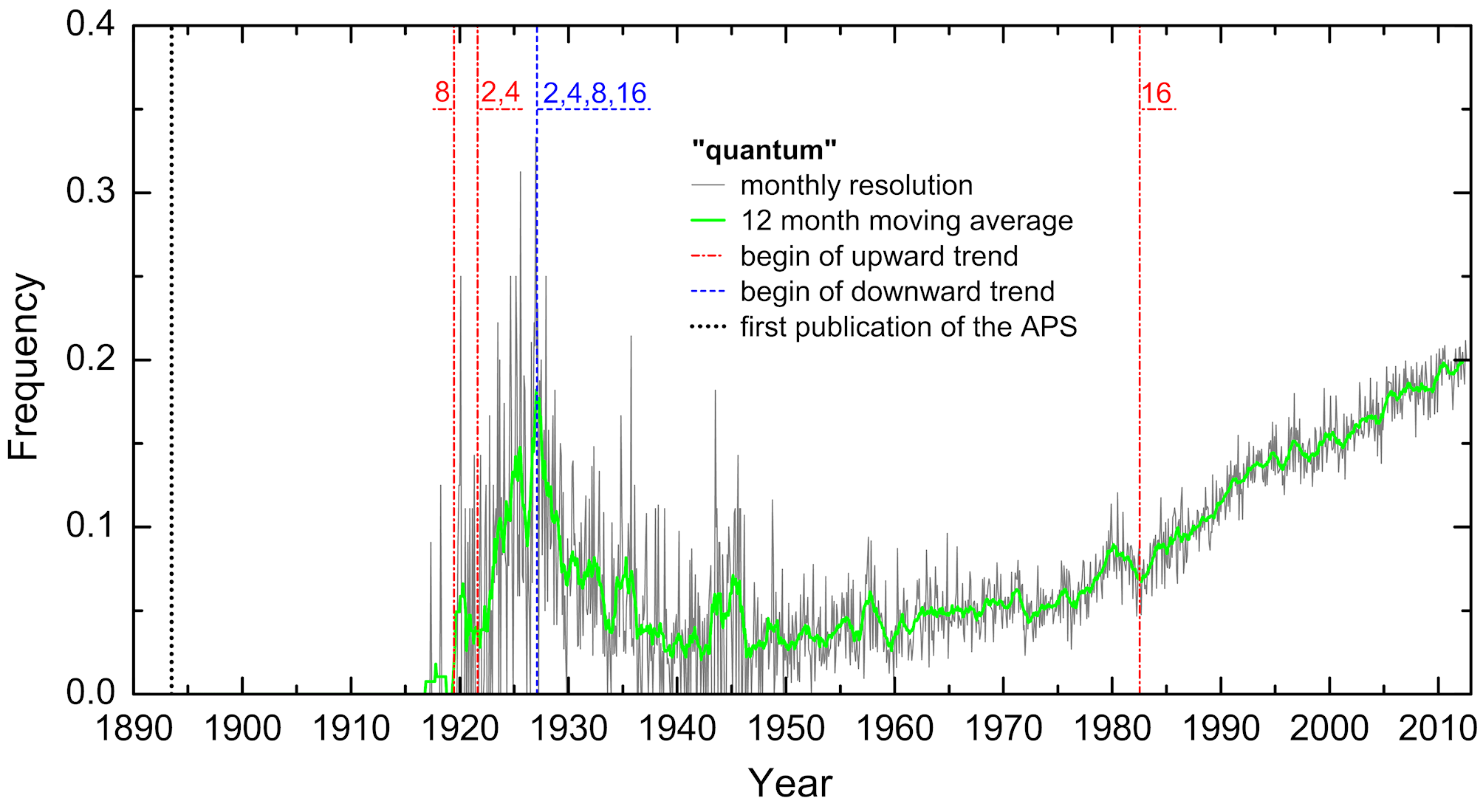}}
\caption{Culturomics of physics enables a quantitative analysis of the trends of scientific discovery. The Physical Review published its first publication in July 1893. Almost 24 years later, in May 1917, the word ``quantum'' is first mentioned in the title of the paper by E. M. Terry, \textit{Phys. Rev.} \textbf{9}, 393. The popularity of the word rises fast and peaks in January 1927 at $f=0.33$, but then starts declining almost as fast as it rose. An upward momentum is picked up again during the 70s, which continues till today. By quantifying the upward and downward trends by means of piecewise linear fitting of the moving average, we can identify the starting points of periods of duration $w=2$, $4$, $8$ and $16$ years during which the trends were the strongest.}
\label{time}
\end{figure}

To do so, we employ the methodology described in the Methods section. Results for the word ``quantum'' are presented in Fig.~\ref{time} (the n-gram viewer for publications of the American Physical Society is available at \href{http://www.matjazperc.com/aps}{\textcolor{blue}{matjazperc.com/aps}}), where the vertical lines denote the starting times of windows during which the maximal upward and downward trends were recorded. By performing the same analysis on the whole data set and ranking the trends in a decreasing manner (we use absolute values for negative slopes $x$), we arrive at the biggest ever upward and downward movers across the whole publishing history of the American Physical Society. Since the obtained tables are too big to be displayed meaningfully in print, we make them available online at \href{http://www.matjazperc.com/aps/rankings}{\textcolor{blue}{matjazperc.com/aps/rankings}}, separately for all time windows $w$ and eligible journals. The Physical Review ST: Physics Education Research and Physical Review X (PRX) do not have an extensive enough publication history to qualify for this analysis. Although it would be interesting to comment on the trends of particular words and phrases and reconcile them with other historical accounts, the options for how to do that are simply too many to be meaningfully covered in this publication. We hope readers will find their favorites amongst the trendsetters and conduct experiments of their own. Here we proceed with the focus on the large-scale properties of the trends.

\begin{figure}
\centering{\includegraphics[width = 8.5cm]{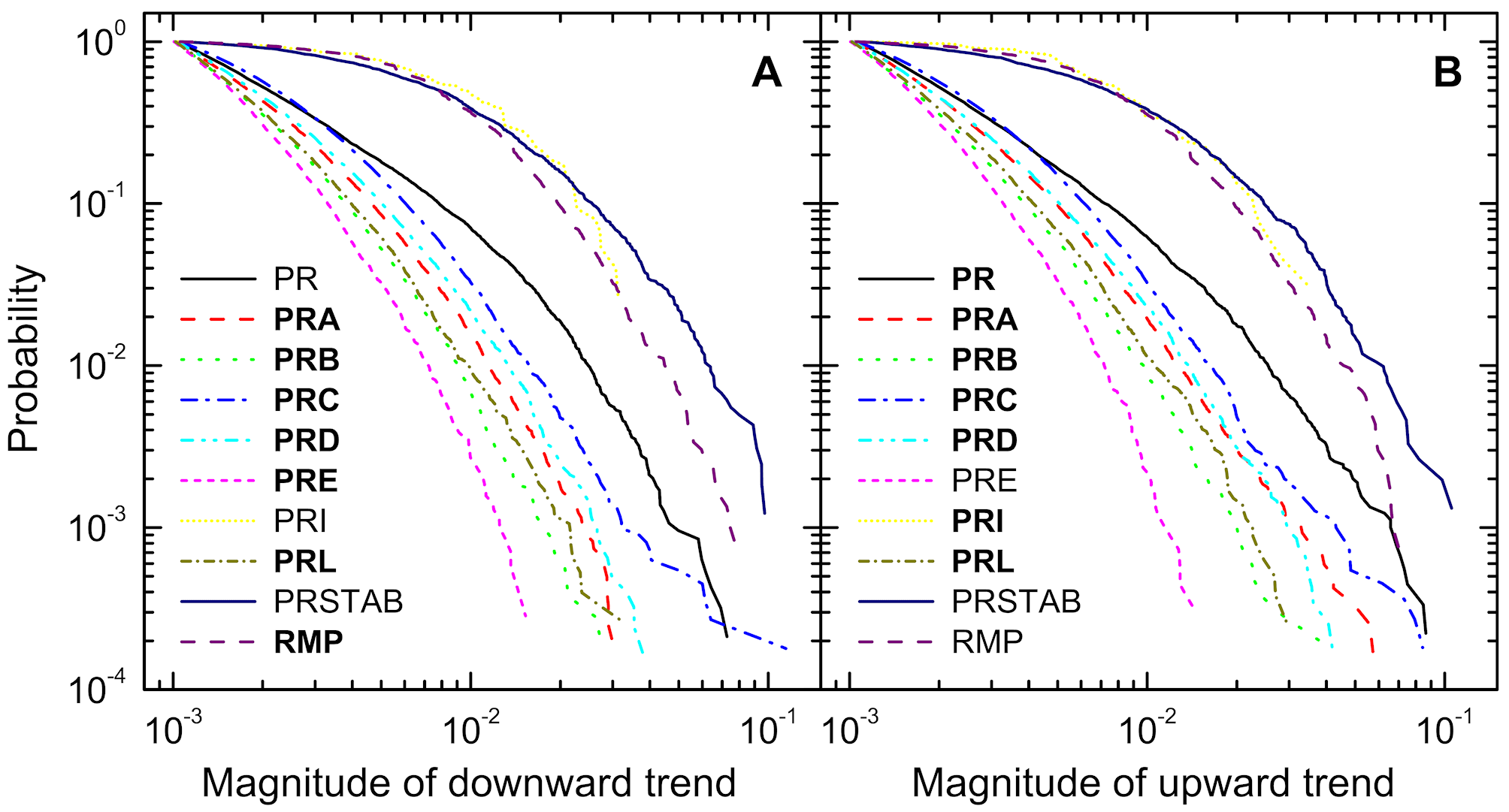}}
\caption{Heavy tails in the distributions of upward and downward trends indicate that the trendsetters in physics are few but strong. The large majority of words and phrases will never reach widespread use. Depicted is the probability that the magnitude of the downward (\textbf{A}) and upward (\textbf{B}) trend will be at least that displayed horizontally. For journals denoted bold the power law gives an acceptable fit to the data (see Table~\ref{fit} for details).}
\label{dist}
\end{figure}

As Fig.~\ref{dist} demonstrates, the distributions of magnitude have heavy tails, largely independent of the direction of trend and journal. Nevertheless, subtle differences can be inferred, and they deserve special attention. To determine the properties of the depicted distributions more accurately, we test several hypotheses. The first is that the depicted cumulative distributions follow a power law $P(x) \propto (x/x_{\min})^{-\alpha+1}$. By using maximum-likelihood fitting methods and goodness-of-fit tests based on the Kolmogorov-Smirnov statistics \cite{clauset_siam09}, we find that only for the journals depicted bold in the legend of Fig.~\ref{dist} the power law is an acceptable fit. The distributions of downward trends for the Physical Review (PR) and Physical Review Series I (PRI) are best described by a power law with an exponential cut-off $P(x) \propto x^{-\alpha+1}\exp(-\lambda x)$, while the distribution of upward trends for the Physical Review E (PRE) is a stretched exponential $P(x) \propto x^{\beta-1}\exp(-\lambda x^{\beta})$. The pertaining exponents are summarized in Table~\ref{fit}.

\begin{table}[b]
\caption{\label{fit} Characterization of the distribution of the magnitude of trends reveals that for the majority the power law is an acceptable hypothesis. Journals for which both the upward and downward trends follow a power law are denoted bold. We use $p>0.1$ as the threshold for acceptance \cite{clauset_siam09}. The distributions of downward trends for PR and PRI are described best by a power law with an exponential cut-off having ($\alpha=1.91$, $\lambda=69.1$) and ($\alpha=0.88$, $\lambda=113.2$), respectively, while the distribution of upward trends for PRE is a stretched exponential with ($\beta=0.62$, $\lambda=121.3$). For the distribution of upward trends for the Reviews of Modern Physics (RMP) and for both distributions concerning Physical Review ST: Accelerators and Beams (PRSTAB), none of the five considered functions, including the exponential and log-normal in addition to the three already mentioned, provide an acceptable fit.}
\begin{center}\begin{tabular}{ c | c | c }\hline
journal & trend & power-law parameters and the goodness-of-fit\\\hline\hline
\multirow{2}{*}{PR}
& $\Uparrow$   & $x_{\min}=0.0212$, $\alpha=3.68$, $p=0.87$\\
& $\Downarrow$ & $x_{\min}=0.0023$, $\alpha=2.29$, $p=0.00$\\\hline
\multirow{2}{*}{\textbf{PRA}}
& $\Uparrow$   & $x_{\min}=0.0061$, $\alpha=3.49$, $p=0.69$\\
& $\Downarrow$ & $x_{\min}=0.0089$, $\alpha=4.17$, $p=0.18$\\\hline
\multirow{2}{*}{\textbf{PRB}}
& $\Uparrow$   & $x_{\min}=0.0052$, $\alpha=3.84$, $p=0.85$\\
& $\Downarrow$ & $x_{\min}=0.0082$, $\alpha=4.42$, $p=0.15$\\\hline
\multirow{2}{*}{\textbf{PRC}}
& $\Uparrow$   & $x_{\min}=0.0084$, $\alpha=3.58$, $p=0.67$\\
& $\Downarrow$ & $x_{\min}=0.0103$, $\alpha=3.73$, $p=0.51$\\\hline
\multirow{2}{*}{\textbf{PRD}}
& $\Uparrow$   & $x_{\min}=0.0116$, $\alpha=4.11$, $p=0.68$\\
& $\Downarrow$ & $x_{\min}=0.0150$, $\alpha=4.77$, $p=0.14$\\\hline
\multirow{2}{*}{PRE}
& $\Uparrow$   & $x_{\min}=0.0026$, $\alpha=3.71$, $p=0.02$\\
& $\Downarrow$ & $x_{\min}=0.0031$, $\alpha=3.89$, $p=0.16$\\\hline
\multirow{2}{*}{PRI}
& $\Uparrow$   & $x_{\min}=0.0079$, $\alpha=2.75$, $p=0.19$\\
& $\Downarrow$ & $x_{\min}=0.0115$, $\alpha=3.21$, $p=0.03$\\\hline
\multirow{2}{*}{\textbf{PRL}}
& $\Uparrow$   & $x_{\min}=0.0029$, $\alpha=3.14$, $p=0.11$\\
& $\Downarrow$ & $x_{\min}=0.0054$, $\alpha=3.87$, $p=0.98$\\\hline
\multirow{2}{*}{PRSTAB}
& $\Uparrow$   & $x_{\min}=0.0144$, $\alpha=2.91$, $p=0.01$\\
& $\Downarrow$ & $x_{\min}=0.0212$, $\alpha=3.28$, $p=0.06$\\\hline
\multirow{2}{*}{RMP}
& $\Uparrow$   & $x_{\min}=0.0231$, $\alpha=4.01$, $p=0.09$\\
& $\Downarrow$ & $x_{\min}=0.0308$, $\alpha=4.46$, $p=0.12$\\\hline
\end{tabular}\end{center}\end{table}

\begin{figure}
\centering{\includegraphics[width = 8.5cm]{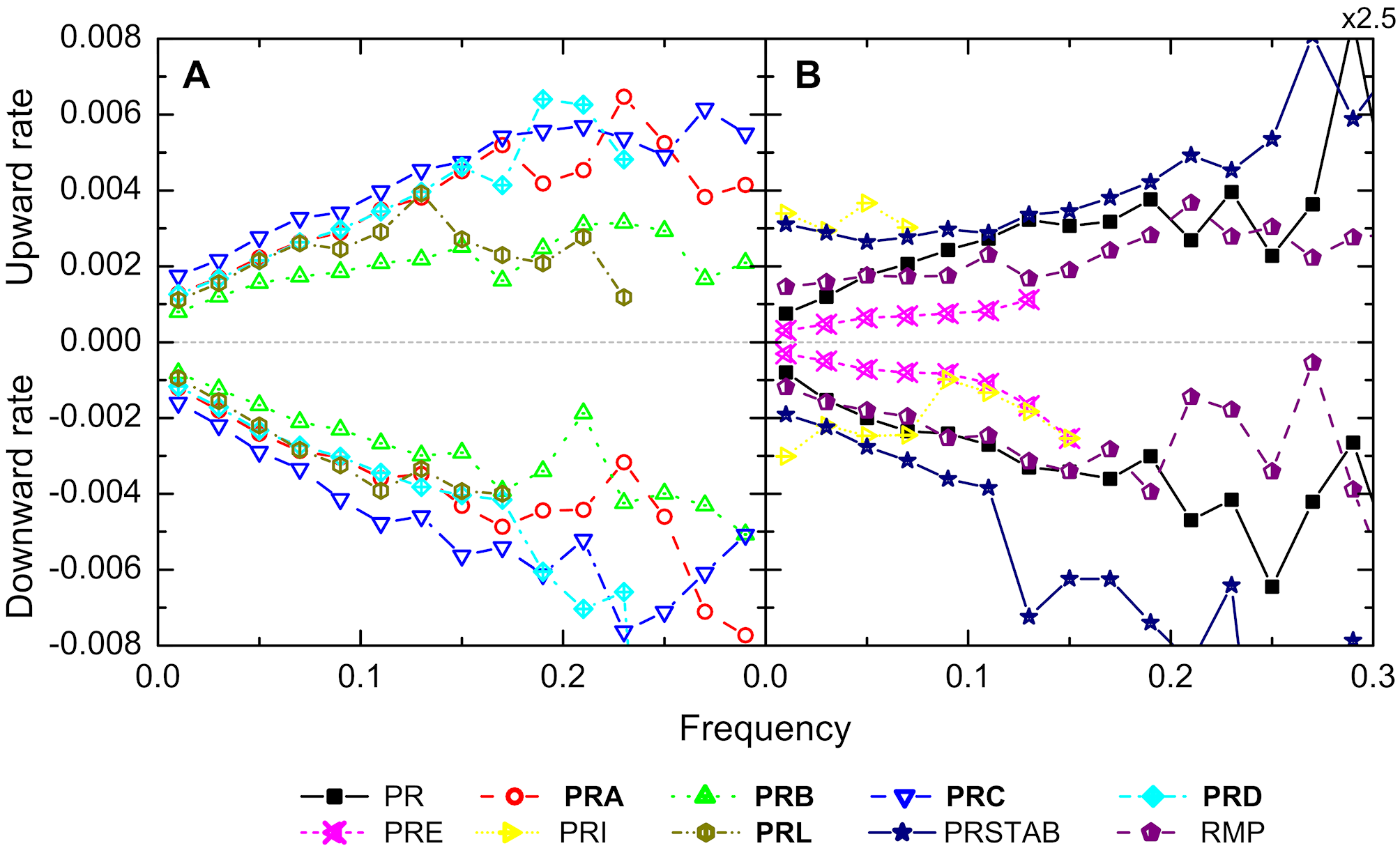}}
\caption{Linear upward and downward rates indicate that the rise and fall of scientific paradigms is governed by robust principles of self-organization. The more a topic is popular, the more popular it is likely to become. Vice versa, during times of decline the fall is going to be the stronger the larger the popularity. (\textbf{A}) Rates per year averaged over top $1000$ words and phrases for journals denoted bold, for which the distribution of both upward and downward trends follows a power law. Beyond $f \approx 0.1$ saturation effects give rise to stronger deviations from the linear form. (\textbf{B}) Rates per year for the remaining journals, for which either the upward or downward trends or both are not satisfactory described by a power law. Deviations from linear rates translate directly to deviations from the power law in the corresponding cumulative distributions depicted in Fig.~\ref{dist}.}
\label{rate}
\end{figure}

Although distributions depicted in Fig.~\ref{dist} are not the most beautiful power laws, and some altogether fail to conform to the power-law hypothesis, the prevalence of heavy tails nevertheless hints firmly towards robust large-scale self-organization governing the up and down trends. By defining the trend rate as $r(f)=\frac{\Delta f(\Delta t)}{\Delta t}$, where $\Delta t$ is the smallest time interval between two consecutive trajectory points, we can directly test for the Matthew effect. However, since the trajectories exhibit both up and down trends, we determine the upward and downward rates separately within time windows of maximal growth and decline. Results presented in Fig.~\ref{rate} confirm that the more commonly used a given word or a phrase is, the larger its expected upward momentum is going to be. The same holds true for the magnitude of falls during times of decline. Together with the continuity of scientific progress, the Matthew effect gives rise to strong but rare trendsetters on the expense of the majority of discoveries that remain forever unknown except to those that made it.

\section*{Discussion}
\noindent The model of growth and preferential attachment \cite{barabasi_s99} captures the essence of our observations. Attachment rates that are linearly proportional with the degree of each node translate into power-law distributions, while deviations from the linear form lead to deviations in the corresponding distributions. Near-linear attachment rates yield log-normal distributions \cite{redner_pt05}, while sublinear attachment rates yield distributions with an exponential cut-off or stretched exponential distributions \cite{dorogovtsev_prl00, krapivsky_prl00}, depending further on the details of sublinearity. The accuracy of empirical studies will also be impaired by finite-size effects and saturation, which may additionally contribute to deviations from the power law \cite{barabasi_pa02}. By contrasting the distributions in Fig.~\ref{dist} with the corresponding rates depicted in Fig.~\ref{rate}, we find an agreement that is well aligned with the theoretical expectations. Moreover, having a closer look at the journals for which the deviations from the linear rates are particularly strong, we find either that they were published in a time when abstracts were rare (PRI, partly also PR), that their publication history is relatively short (PRE, PRSTAB), or that they publish reviews rather than original research (RMP), all of which are probable causes for the analysis on this particular cases to give less conclusive results.

The identified self-organization in the rise and fall of scientific paradigms can be seen akin to previous reports of preferential attachment in citation rates and the acquisition of scientific collaborators \cite{redner_pt05, newman_pre01}. Specifically related to the former case, our discovery can be interpreted as the textual extension of the Matthew effect in citation rates or as the large-scale ``semantic'' version of that effect. It is also worth noting that, although it is debatable whether the concept of preferential attachment is based on luck or reason \cite{barabasi_n12}, in our case at least it seems inevitably due to the actual progress made, not chance that could make one discovery seem bigger than it truly is.

\section*{Methods}
\noindent After identifying all unique words and phrases, we determine their relative frequency of occurrence $f$ with respect to the number of publications in any given month for each journal published by the American Physical Society as well as overall. We consider a phrase to be a string of words separated by a space, and we limit our analysis to at most four-word phrases to keep the volume of information to be processed manageable. By ignoring capitalization, numbers, words containing numbers, and formulae, we identify 118056 single words, 3269090 two-word phrases, 13295156 three-word phrases, and 23799449 four-word phrases, thus obtaining over 40 million trajectories that enable a qualitative exploration of the trends of physics discovery. While of course not all identified words and phrases have to do with physics, the assumption we make is that only those that do will actually exhibit notable trends. Words like ``the'' or ``of'' appear in nearly every abstract. The word ``quantum'', on the other hand, is mentioned first in the 1917 May issue of the Physical Review, with popularity subsequently peaking in January 1927 at $f=0.33$, as depicted in Fig.~\ref{time}. Trajectories of all other words and phrases can be searched for and viewed at \href{http://www.matjazperc.com/aps}{\textcolor{blue}{matjazperc.com/aps}}.

However, since not all publications of the American Physical Society have an abstract, and since some abstracts are very short, even the most common words and phrases can occasionally exhibit relatively strong trends. Not to treat those trends as trends of physics discovery, we eliminate from the analysis the most common English words as identified in \cite{perc_jrsi12}, minus a few hand-picked special cases that obviously have to do with physics. We also eliminate phrases that contain the most common English words either at the beginning or end. With these two additional filters in place, we make sure that from all the identified unique words and phrases the focus is on those that, in the majority of cases, concern at least some aspect of physics.

To quantify the trends, we seek out time windows where the slope $x$ of the linear fit of each trajectory is maximally positive and maximally negative, and we do so separately for windows of width $w=2$, $4$, $8$ and $16$ years. The dispersal in years is important as we want our analysis to encompass short-, mid- and long-term trends. Although a straight line won't be a good fit for the data in several cases, it is nevertheless a useful first-order approximation for whether a subject is trending up or down, and to what extent this is the case \cite{akritas_jass95}. Recent most advances on how to identify trends in word frequency dynamics are presented in \cite{altmann_jstat13}, and they shall be an excellent basis for future explorations. As starting points of each of the four considered time windows, we consider every month of every word and phrase for which data is available, with the obvious condition that the starting point plus the window width must not go beyond October 2012. Before the analysis we apply a 12 month moving average on the trajectories and require that the considered time windows must not contain missing data after averaging. Moreover, we dismiss all words and phrases with $\max|x|<0.001$/year as lacking notable trends.

\begin{acknowledgments}
\noindent We thank the American Physical Society for granting permission to use the data set and Mark Doyle for providing it. This research was supported by the Slovenian Research Agency (Grant J1-4055).
\end{acknowledgments}

\end{document}